\documentclass[
  10pt,
  twocolumn,
  twoside,
]{article}

\usepackage[numbers,round,comma,sort&compress]{natbib}

\usepackage[
  left=0.65in,
  right=0.65in,
  top=0.65in,
  bottom=0.65in,
]{geometry}

\usepackage{titlesec}
\titleformat{\section}{\raggedright\large\bfseries}{\thesection}{1em}{}
\titleformat{\subsection}{\raggedright\large}{\thesubsection}{1em}{}
\usepackage{ragged2e}

\usepackage[font=small,labelfont=bf]{caption}

\usepackage{fancyhdr}
\usepackage[realmainfile]{currfile}
\input{\currfilebase.settings}
\usepackage{environ}

\usepackage{fancyhdr}
\usepackage[export]{adjustbox}

\usepackage{colortbl}

\PassOptionsToPackage{hyphens}{url}\usepackage{hyperref}
\makeatletter
\g@addto@macro{\UrlBreaks}{\UrlOrds}
\makeatother

\hypersetup{
  colorlinks=true,
  allcolors=todoblue,
  urlcolor=todoblue,
  citecolor=todoblue,
  pdfborder={0 0 0},
  breaklinks=true,
}

\usepackage[normalem]{ulem}

\usepackage{soul}

\makeatletter
\let\ftype@table\ftype@figure
\makeatother

\usepackage{makecell}

\usepackage{array}

\usepackage{xcolor}

\definecolor{todoblue}{RGB}{0, 91, 187}

\usepackage{tocloft}

\newenvironment{textblock}{\renewcommand{\item}{}\ignorespaces}{}

\usepackage{changepage}

\usepackage{graphicx}
\usepackage{verbatim}
\usepackage{amssymb}
\usepackage{amsmath}
\usepackage{ifthen}

\usepackage{longtable}

\usepackage{xcolor}

\definecolor{olivegreen}{rgb}{0.33333,.41961,0.18431}
\definecolor{forestgreen}{rgb}{0.13333,.5451,0.13333}

\definecolor{lightgrey}{rgb}{0.7,0.7,0.7}
\definecolor{verylightgrey}{rgb}{0.90,0.90,0.90}
\definecolor{veryverylightgrey}{rgb}{0.95,0.95,0.95}
\definecolor{grey}{rgb}{0.5,0.5,0.5}
\definecolor{darkgrey}{rgb}{0.3,0.3,0.3}
\definecolor{verydarkgrey}{rgb}{0.15,0.15,0.15}

\definecolor{headerblue}{HTML}{33367E}
\definecolor{unitednationsblue}{HTML}{4D88FF}

\definecolor{charcoal}{HTML}{36454F}
\definecolor{cinerous}{HTML}{98817B}
\definecolor{feldgrau}{HTML}{4D5D53}
\definecolor{glaucous}{HTML}{6082B6}
\definecolor{arsenic}{HTML}{3B444B}
\definecolor{xanadu}{HTML}{738678}

\definecolor{firebrick}{HTML}{B22222}
\definecolor{orangered}{HTML}{FF4500}
\definecolor{tomato}{HTML}{FF6347}

\definecolor{orange}{RGB}{255,116,0}

\definecolor{purpletaupe}{HTML}{3B444B}

\definecolor{rose}{HTML}{E3242B}

\colorlet{editnotecolor}{rose}

\definecolor{headerorange}{RGB}{255,116,0}
\definecolor{headergray}{RGB}{230,230,230}

\definecolor{headerpop}{RGB}{230,230,230}

\definecolor{magmalight}{RGB}{252,251,195}
\definecolor{magmalightalt}{RGB}{250,240,184}
\definecolor{magmamedium}{RGB}{245,200,146}
\definecolor{magmadark}{RGB}{224,106,98}

\definecolor{icelight}{RGB}{223,242,244}
\definecolor{icelightalt}{RGB}{189,222,226}
\definecolor{icemedium}{RGB}{132,184,204}
\definecolor{icedark}{RGB}{103,153,191}

\definecolor{traitrowcolor}{RGB}{223,242,244}
\definecolor{traitrowcoloralt}{RGB}{189,222,226}

\definecolor{characterrowcolor}{RGB}{252,251,195}
\definecolor{characterrowcoloralt}{RGB}{250,240,184}

\definecolor{archetyperowcolor}{RGB}{255,213,212} 
\definecolor{archetyperowcoloralt}{RGB}{255,182,179} 

\definecolor{datasetrowcolor}{RGB}{232,244,234}
\definecolor{datasetrowcoloralt}{RGB}{210,231,214}

\NewEnviron{excerpt}{
    \begin{quote}
      \medskip
      \BODY
      \medskip
    \end{quote}
}

\NewEnviron{editnote}{
    \begin{quote}
      \color{rose}
      $\blacksquare$
      \BODY
      \medskip
    \end{quote}
}

\newcommand{\command}[1]{
  \lstinline[language={[LaTeX]TeX},basicstyle=\ttfamily]{#1}
}

\newcommand{\editbox}[2]{
}

\newcommand{\editboxwithlatex}[2]{
}

\usepackage{fancyvrb}

\usepackage{tcolorbox}
\tcbuselibrary{skins,breakable,listings,breakable}

\usepackage{tikz}
\usetikzlibrary{shapes}

\tikzstyle{mybox} = [draw=lightblue!70, fill=lightblue!7, very thick,
    rectangle, rounded corners, inner sep=10pt, inner ysep=20pt]

\tikzstyle{editortitle} =[draw=archetyperowcoloralt, fill=archetyperowcoloralt, text=black]

\newcommand\Loadedframemethod{default}
\usepackage[framemethod=\Loadedframemethod]{mdframed}

\mdfsetup{skipabove=\topskip,skipbelow=\topskip}

\tikzstyle{loglinetitle} =[draw=icedark, fill=icemedium!50, text=black]

\newenvironment{loglinebox}[1][]{

  \ifstrempty{#1}%
  {\mdfsetup{%
    frametitle={%
       \tikz[baseline=(current bounding box.east),outer sep=0pt]
        \node[loglinetitle, anchor=east,rectangle]
        {\strut~~#1:~~\strut};}}
  }%
  {\mdfsetup{%
     frametitle={%
       \tikz[baseline=(current bounding box.east),outer sep=0pt]
        \node[loglinetitle,anchor=east,rectangle]
        {\strut~~#1:~~\strut};}}%
   }%
   \mdfsetup{innertopmargin=5pt,linecolor=icedark,%
             linewidth=0.5pt,topline=true,
             frametitleaboveskip=\dimexpr-\ht\strutbox\relax,}
   \begin{mdframed}[backgroundcolor=icelight,nobreak=true]\relax%
     \raggedright
}{\end{mdframed}}

\tikzstyle{abstracttitle} =[draw=magmadark!75, fill=magmamedium!75, text=black]

\newenvironment{abstractbox}[1][]{

  \ifstrempty{#1}%
  {\mdfsetup{%
    frametitle={%
       \tikz[baseline=(current bounding box.east),outer sep=0pt]
        \node[abstracttitle, anchor=east,rectangle]
        {\strut~~#1:~~\strut};}}
  }%
  {\mdfsetup{%
     frametitle={%
       \tikz[baseline=(current bounding box.east),outer sep=0pt]
        \node[abstracttitle,anchor=east,rectangle]
        {\strut~~#1:~~\strut};}}%
   }%
   \mdfsetup{innertopmargin=5pt,linecolor=magmadark,%
             linewidth=0.5pt,topline=true,
             frametitleaboveskip=\dimexpr-\ht\strutbox\relax,}
   \begin{mdframed}[backgroundcolor=magmalight,nobreak=true]\relax%
     \raggedright
}{\end{mdframed}}

\tikzstyle{infotitle} =[draw=darkgrey, fill=lightgrey!50, text=black]

\newenvironment{infobox}[1][]{

  \ifstrempty{#1}%
  {\mdfsetup{%
    frametitle={%
       \tikz[baseline=(current bounding box.east),outer sep=0pt]
        \node[infotitle, anchor=east,rectangle]
        {\strut~~#1:~~\strut};}}
  }%
  {\mdfsetup{%
     frametitle={%
       \tikz[baseline=(current bounding box.east),outer sep=0pt]
        \node[infotitle,anchor=east,rectangle]
        {\strut~~#1:~~\strut};}}%
   }%
   \mdfsetup{innertopmargin=5pt,linecolor=grey,%
             linewidth=0.5pt,topline=true,
             frametitleaboveskip=\dimexpr-\ht\strutbox\relax,}
   \begin{mdframed}[backgroundcolor=lightgrey!25,nobreak=true]\relax%
     \raggedright
}{\end{mdframed}}

\usepackage{enumitem}

\tikzstyle{changelogtitle} =[draw=darkgrey, fill=lightgrey!50, text=black]

\usepackage{listings}

\usepackage{stringstrings}
\usepackage{readarray}

\def\firstchar#1#2|{#1}
\edef\tbs{\detokenize{\X}}
\edef\tbs{\expandafter\firstchar\tbs|}
\edef\tlb{\detokenize{{}}}
\edef\tlb{\expandafter\firstchar\tlb|}
\edef\tus{\detokenize{_}}
\newcounter{tokenindex}
\newcommand\detokenizeplus[1]{%
  \def\temparg{\detokenize{#1}}%
  \getargsC{\temparg}%
  \setcounter{tokenindex}{0}%
  \def\prevmacro{F}%
  \whiledo{\value{tokenindex} < \narg}{%
    \stepcounter{tokenindex}%
    \isnextbyte[q]{\tbs}{\csname arg\roman{tokenindex}\endcsname}%
    \if T\theresult%
      \if T\prevmacro\unskip\else\fi%
      \def\prevmacro{T}%
    \else%
      \def\prevmacro{F}%
   \fi%
    \isnextbyte[q]{\tlb}{\csname arg\roman{tokenindex}\endcsname}%
    \if T\theresult\unskip\else\fi%
    \isnextbyte[q]{\tus}{\csname arg\roman{tokenindex}\endcsname}%
    \if T\theresult\unskip\else\fi%
    \csname arg\roman{tokenindex}\endcsname~%
  }%
}

\usepackage{mathtools}

\newboolean{twocolswitch}

\newcommand{\sindex}[1]{}
\newcommand{\nindex}[1]{}

\newcommand{\www}[1]{\url{#1}}

\newcommand{\Req}[1]{Eq.~(\ref{#1})}

\usepackage{lettrine}

\usepackage{etoolbox,refcount}

\newcounter{countitems}
\newcounter{nextenumeratecount}
\newcommand{\setupcountitems}{%
  \stepcounter{nextenumeratecount}%
  \setcounter{countitems}{0}%
  \preto\item{\stepcounter{countitems}}%
}
\makeatletter
\newcommand{\computecountitems}{%
  \edef\@currentlabel{\number\c@countitems}%
  \label{countitems@\number\numexpr\value{nextenumeratecount}-1\relax}%
}
\newcommand{\nextenumeratecount}{%
  \getrefnumber{countitems@\number\c@nextenumeratecount}%
}
\makeatother
\AtBeginEnvironment{enumerate}{\setupcountitems}
\AtEndEnvironment{enumerate}{\computecountitems}

\usepackage{xfrac}

\newcommand{\dee}[1]{\textnormal{d}#1}

\newcommand{\diff}[2]{\frac{{\rm d}#1}{{\rm d}#2}}

\newcommand{\rank}{r}
\newcommand{\rgrtime}{t}
\newcommand{\dummytime}{\rgrtime'}
\newcommand{\rgrindex}{i}

\newcommand{\typeinittime}{\rgrtime_{\rank}^{\textnormal{init}}}
\newcommand{\typeinittimefn}[1]{\rgrtime_{#1}^{\textnormal{init}}}

\newcommand{\sizerankingexponent}{\alpha}

\newcommand{\innovationprob}{\rho}
\newcommand{\innovationrate}{\innovationprob_{\rgrtime,\alpha}}
\newcommand{\innovationratefn}[2]{\innovationprob_{#1,#2}}

\newcommand{\typesize}{S}

\newcommand{\growthfactorsymbol}{G}
\newcommand{\growthfactor}{\growthfactorsymbol_{\rgrtime,\sizerankingexponent}}

\newcommand{\Ndistincttypes}{\numbersymbol_{\rgrtime,\sizerankingexponent}}
\newcommand{\Ndistincttypesfn}[2]{\numbersymbol_{#1,#2}}

\newcommand{\rgrconstant}{g}

\newcommand{\groupsize}{k}

\newcommand{\numbersymbol}{N}

\newcommand{\bigprobsymbol}{P}

\setboolean{twocolswitch}{true}

\usepackage{natbib}
\setcitestyle{square}
\raggedright

\newcommand{\zenodorecord}{15006457}
\newcommand{\zenodourl}{https://doi.org/10.5281/zenodo.\zenodorecord}

\newcommand{\gitlaburl}{https://gitlab.com/compstorylab/allotaxonometer/}

\newcommand{\onlineappendicesurl}{https://compstorylab.org/allotaxonometry/}

\usepackage{authblk}

\begin{document}

\title{
  \textbf{\protect

Simon's model does not produce Zipf's law: 
\\
The fundamental rich-get-richer mechanism \\
for any power-law size ranking

}
}


\renewcommand*{\Authsep}{, }
\renewcommand*{\Authand}{, }
\renewcommand*{\Authands}{, }
\renewcommand*{\Affilfont}{\normalsize\normalfont}
\renewcommand*{\Authfont}{}
\setlength{\affilsep}{2em}

\author[1,2]{Pablo~Rosillo-Rodes}

\author[2,3,4]{Julia~Witte~Zimmerman}


\author[4,5]{Laurent~H{\'e}bert-Dufresne}

\author[4,5,6]{Peter~Sheridan~Dodds}


\affil[1]{
 Institute for Cross-Disciplinary Physics and Complex Systems IFISC (UIB-CSIC), Campus Universitat de les Illes Balears, E-07122 Palma de Mallorca, Spain
}

\affil[2]{
  Computational Story Lab,
  University of Vermont,
  Burlington,
  VT 05405,
  US
}

\affil[3]{
  Computational Ethics Lab,
  University of Vermont,
  Burlington,
  VT 05405,
  US
}

\affil[4]{
  Vermont Complex Systems Institute,
  Vermont Advanced Computing Center,
  MassMutual Center of Excellence for Complex Systems and Data Science,
  University of Vermont,
  Burlington,
  VT 05405,
  US
}

\affil[4]{
  Department of Computer Science,
  University of Vermont,
  Burlington,
  VT 05405,
  US
}

\affil[5]{
  Santa Fe Institute,
  1399 Hyde Park Rd,
  Santa Fe,
  NM 87501,
  US
}

\affil[6]{
  Complexity Science Hub,
  Metternichgasse 8,
  1030 Vienna,
  Austria
}

\date{\today}
\maketitle

\begin{center}
  \begin{minipage}[t]{0.85\linewidth}
    \begin{loglinebox}[Depiction]
      \centering

\begin{quote}
    Jesus said good-bye and departed, his host's parting words ringing in his ears, Blessed be You, O Lord our God, King of the Universe, who guides our footsteps, words he repeated to himself, praising that same Lord, God, and King, provider of all our needs, as can clearly be seen from everyday experience, in accordance with that most just rule of direct proportion, which says that more should be given to those who have more.
\end{quote}
\bigskip
{\raggedleft
--- José Saramago 
\\
\textit{The Gospel according to Jesus Christ}
}
    \end{loglinebox}
  \end{minipage}\hfill
  \begin{minipage}[t]{\linewidth}
    \centering
    \includegraphics[width=\linewidth]{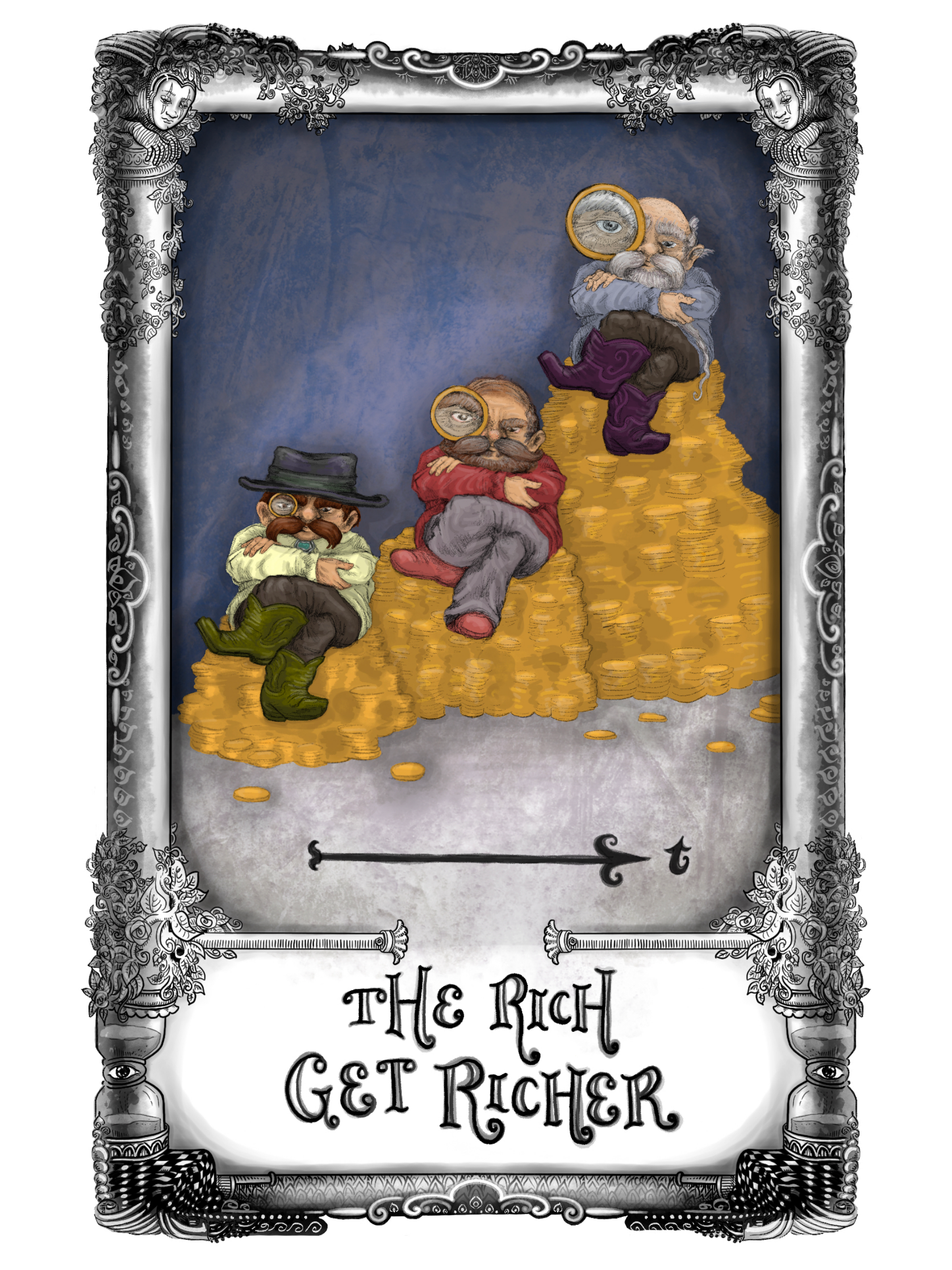}
    \label{fig:tarot_card_richer}
  \end{minipage}
\end{center}

\clearpage

\newgeometry{
  left=2in,
  right=2in,
  top=1in,
  bottom=1in,
  }

\onecolumn

\renewcommand{\baselinestretch}{1.25}
\selectfont


\begin{abstractbox}[Logline]
  \raggedright
  \begin{textblock}
\item Power-law size rankings---from word frequencies to city sizes---are among the most universal patterns in complex systems, yet their canonical model, due to Simon, catastrophically fails in the Zipf's law limit.
\item 
We derive the fundamental rich-get-richer mechanism that correctly produces any power-law size ranking, uncovering that Zipf's law emerges when innovation decays as the inverse logarithm of the number of types.
\item 
Our dynamic innovation rate universally governs type emergence in any such system, regardless of the underlying mechanism.
\end{textblock}

  \smallskip
\end{abstractbox}


\begin{abstractbox}[Abstract]
  \raggedright
  \begin{textblock}
\item
  Many complex systems
  are composed of disparate, interacting types of
  varying `sizes':
  Species abundances in ecosystems,
  firm sizes in markets,
  city populations in countries,
  word counts in language, etc.
\item 
  A longstanding mystery of complex systems is Zipf's law,
  which is the empirical observation that
  component size
  decreases as the inverse of component rank---$\typesize \propto \rank^{-1}$--- and its generalization 
  $\typesize \propto \rank^{-\sizerankingexponent}$
  for 
  $\sizerankingexponent \ge 0$.
\item 
  Herbert Simon's 1955 theoretical rich-get-richer mechanism
  for system growth has prevailed
  as capturing the essential process.
\item
  But Simon's analysis is in fact flawed:
  In the limit of zero innovation,
  the model leads to a winner-takes-all system
  with $\sizerankingexponent \rightarrow \infty$,
  rather than $\sizerankingexponent \rightarrow 1$.
\item
  Here,
  for pure rich-get-richer systems,
  we derive 
  the time-dependent innovation rate 
  $\innovationprob_{\rgrtime}$
  that
  correctly produces power-law size rankings across
  all $\sizerankingexponent \ge 0$.
\item
  To produce Zipf's law,
  we uncover that 
  $\innovationprob_{\rgrtime}$
  must decay as the inverse of 
  the log of the number of types,
  $1/\ln N$. 
\item
  We then show that our time-dependent innovation rate governs
  type emergence in any system obeying a power-law size-ranking,
  independent of the underlying mechanism.
\item 
  We demonstrate agreement between our model's output
  and word rankings in
  a collection of famous novels,
  while Simon's model fails.
\item
  Going forward, our dynamic innovation rate mechanism
  provides the fundamental, Drosophila-like model
  for all rich-get-richer systems.
\end{textblock}

  \smallskip
\end{abstractbox}


\begin{infobox}[Keywords]
  complex systems,
ranks,
size rankings,
power laws, 
scaling,
growth mechanisms,
rich-get-richer, 
innovation,
groups,
language,
ecology,
social sytems,
Zipf, 
Simon, 
types,
tokens,
type-token law, 
finite size
 
 \smallskip
\end{infobox}

\renewcommand{\baselinestretch}{1}
\selectfont

\twocolumn

\restoregeometry

\clearpage



\renewcommand{\floatpagefraction}{0}
\setlength{\parindent}{0pt}

\section{Introduction}
\label{sec:rich-get-richer-complete.introduction}

\begin{textblock}
\item
Power-law size rankings---the empirical observation that component size $\typesize$ decreases as an inverse power of component rank $\rank$, $\typesize \sim \rank^{-\sizerankingexponent}$---are among the most pervasive quantitative regularities in complex systems.
\item
They appear in word frequencies in texts~\cite{estoup1916a,zipf1946,altmann2016,arnon2025}, 
species abundances in ecosystems~\cite{camacho2001}, city populations and firm sizes in socioeconomic systems~\cite{serrano2003,gabaix2016}, 
citation counts in complex networks~\cite{price1965a,Redner1998}, 
and many other contexts~\cite{mitzenmacher2004,newman2005,maillart2008a,roman2022,holehouse2025}.
\item
The special case $\sizerankingexponent = 1$, known as Zipf's law, has attracted particular attention for its near-universality in language and beyond.
\item
Although the empirical universality of power-law size rankings has been questioned in specific domains~\cite{ferrericancho2010a,broido2019,gerlach2019}, the pattern is sufficiently widespread and consistent to demand a principled theory.
\end{textblock}

\begin{textblock}
\item
The most influential explanation is Herbert Simon's 1955 rich-get-richer model~\cite{simon1955a}, which builds on earlier work by Yule~\cite{yule1925a} and has become the canonical theoretical foundation for power-law size rankings.
\item
Simon's mechanism is elegantly simple: at each step, a single new token is added to a growing system, either introducing a novel type with probability $\innovationprob$ or reinforcing an existing type with probability $1-\innovationprob$, where each type attracts new tokens in proportion to its current size.
\item
Simon's rich-get-richer process produces a power-law size ranking with exponent $\sizerankingexponent = 1 - \innovationprob$, and variants of the mechanism have been identified in operation across a wide range of real systems~\cite{merton1988matthewII,Redner1998,maillart2008a}.
\item 
Simon's model was ported to evolving networks by Price for scientific citations in the 1960s~\cite{price1965a}, 
and later separately discovered for so-called scale-free networks~\cite{albert1999}.
\item 
The major competing theory has been Mandelbrot's contemporary optimization argument, which offered no mechanism, did not have a reason for the exceptionality of Zipf's law, and led to an extended back and forth between
Simon and Mandelbrot~\cite{mandelbrot1959a,simon1960a,mandelbrot1961a,simon1961a,mandelbrot1961b,simon1961b}.
\end{textblock}

\begin{figure}[t]
  \includegraphics[width=\columnwidth]{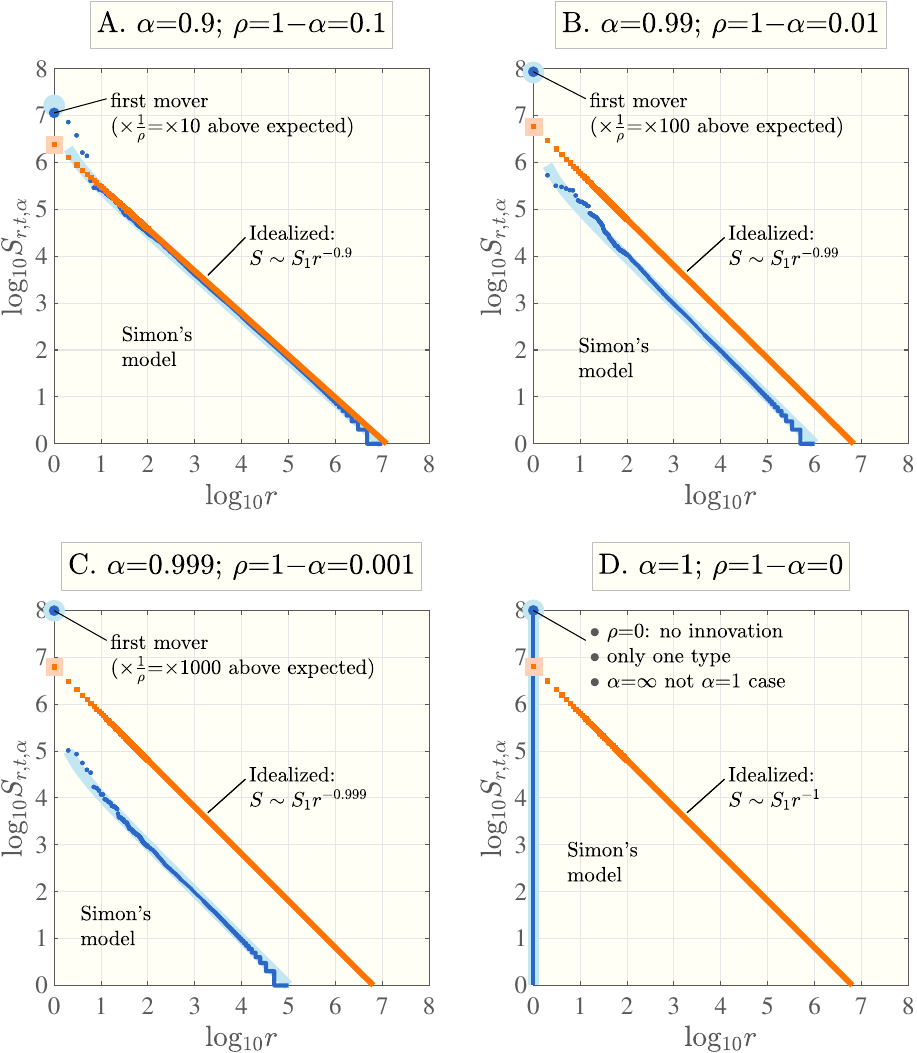}
  \caption{
  Simulations of Simon's model fail to reproduce the size ranking  $\typesize_{\rank,\rgrtime,\sizerankingexponent}$ in the zero innovation limit. As $\innovationprob \to 0$ (\textbf{A}--\textbf{C}), the size ranking fails to approach a power law with $\sizerankingexponent = 1 - \innovationprob \to 1$. Instead, the first-mover advantage dominates and overestimates $\typesize_{0,\rgrtime,\sizerankingexponent}$ in a factor of $1/\innovationprob$. This effect intensifies as the innovation probability decreases, arriving at $\sizerankingexponent \rightarrow \infty$ when $\innovationprob = 0$ (\textbf{D}).
  }
  \label{fig:rich-get-richer-complete.simon-failure}
\end{figure}

\begin{textblock}
\item
But Simon's model contains a long missed critical flaw that becomes catastrophic where it matters most, as $\sizerankingexponent \rightarrow 1^{-}$.
\item
As $\innovationprob \rightarrow 0$, Simon's analysis 
incorrectly showed that the size-ranking exponent approaches $\sizerankingexponent \rightarrow 1^{-}$, recovering Zipf's law.
\item
Instead, the first-mover advantage inherent to rich-get-richer growth diverges as $1/\innovationprob$, so that the leading type's size is a factor of $1/\innovationprob$ larger than all others~\cite{dodds2017pre}.
\item
In the zero-innovation limit $\innovationprob = 0$, the model collapses entirely to a winner-takes-all outcome with the degenerate case of $\sizerankingexponent \rightarrow \infty$ rather than $\sizerankingexponent = 1$.
\item
As we show in Fig.~\ref{fig:rich-get-richer-complete.simon-failure}, this failure intensifies monotonically as $\innovationprob$ decreases.
\item
Furthermore, the model is structurally incapable of producing size rankings with $1 < \sizerankingexponent < \infty$, excluding an entire and empirically relevant region of the parameter space.
\end{textblock}

\begin{textblock}
\item
Here, we derive the correct, time-dependent innovation rate $\innovationrate$ that generalizes Simon's rich-get-richer model to produce power-law size rankings for any $\sizerankingexponent \geq 0$.
\item
Our generalized rate contains no discontinuities but rather
smoothly decreases as the system grows in a manner governed by the target $\sizerankingexponent$ and the current number of distinct types.
\item
For $\sizerankingexponent \ll 1$, the rate reduces to Simon's constant $\innovationprob = 1 - \sizerankingexponent$, recovering his result in the regime where it is correct.
\item 
For $\sizerankingexponent > 1$,
our model again provides a mechanistic rich-get-richer explanation,
which goes beyond an observation of type growth~\cite{Zanette2005}.
\item
For $\sizerankingexponent = 1$, we uncover the innovation rate
behind Zipf's law: 
$\innovationrate \rightarrow 1 / \log \Ndistincttypes$.
\end{textblock}

\begin{textblock}
A broader result is that this dynamic innovation rate arises not only from the mechanistic rich-get-richer process but from any system with a power-law size ranking, independent of the underlying generative process---a consequence of the exact type-token relationship derived in prior work~\cite{rosillo-rodes2026a}.
\item
We validate the generalized model against word size rankings from eight major literary works across six languages, where it succeeds and Simon's model fails.
\item
The generalized innovation rate thereby provides the fundamental reference model for all rich-get-richer phenomena, and a standard against which departures from power-law universality can be measured.
\end{textblock}

\begin{textblock}
\item 
In Section~\ref{sec:rich-get-richer-complete.failure}, we formalize the error in the traditional formulation of Simon's model as an amplification of the first-mover advantage. 
\item 
In Section~\ref{sec:rich-get-richer-complete.generalized-model}, we derive a generalized innovation rate, which corrects the traditional formulation and extends it to include all possible power-law size rankings.
\item 
Finally, in Section~\ref{sec:rich-get-richer-complete.nonmech}, we demonstrate that this generalized innovation rate arises not only from the mechanistic model that we propose here but also characterizes the rate at which the set of types expands in a growing, deterministic, non-mechanistic model.
\item 
In other words, any system described by a power-law size ranking will also adhere to our generalized innovation rate, not just those systems explicable by rich-get-richer processes.
\end{textblock}

\section{The catastrophic failure of Simon's model in the zero innovation limit}
\label{sec:rich-get-richer-complete.failure}

\begin{textblock}
\item
  Simon's analysis~\cite{simon1955a} used a rate equation approach to determine
  the expected number of types of size $\groupsize$ at time $\rgrtime$,
  $\numbersymbol_{\groupsize,\rgrtime,\sizerankingexponent}$.
\item
  The form of
  $\numbersymbol_{\groupsize,\rgrtime,\sizerankingexponent}$
  follows from the recursion relation
\item
  \begin{equation}
    \numbersymbol_{\groupsize,\rgrtime,\sizerankingexponent}
    =
    \frac{
      \groupsize-1
    }{
      \groupsize
      +
      \frac{
        1
      }{
        (1-\innovationprob)
      }
    }
    \numbersymbol_{\groupsize-1,\rgrtime,\sizerankingexponent}
    \label{eq:rich-get-richer-complete.simon's-solution-Nkt-recursive}
  \end{equation}
\item
  for $\groupsize > 1$, where
\end{textblock}
\begin{equation}
  \numbersymbol_{1,\rgrtime,\sizerankingexponent} 
  =
  \frac{
    \innovationprob
  }{
    2 - \innovationprob
  }.
  \label{eq:rich-get-richer-complete.simon's-solution-N1}
\end{equation}

\begin{textblock}
\item
  For large $\groupsize$ and $\rgrtime$, and with the Beta function, $B(\cdot)$, 
  we have
\begin{align}
\notag
  \numbersymbol_{\groupsize,\rgrtime,\sizerankingexponent}
  &\propto
  \frac{
    (\groupsize-1)
    \innovationprob
  }{
    2
    -
    \innovationprob
  }
  B\left(
  \groupsize-1,
  \frac{
    1
  }{
    1-\innovationprob
  }
  +
  2
  \right) \\
  &\sim
  \groupsize^{
    -
    \left(
    1
    +
    \frac{
      1
    }{
      1-\innovationprob
    }
    \right)
  },
  \label{eq:rich-get-richer-complete.simon's-solution-Nkt}
\end{align}
\item
which corresponds to a size ranking given by
\begin{equation}
    \typesize_{\rank,\rgrtime,\sizerankingexponent} \sim \rank^\sizerankingexponent,
    \label{eq:rich-get-richer-complete.simons-tail}
\end{equation}
with $\sizerankingexponent = 1-\innovationprob$. 
Apparently, Simon found that $\sizerankingexponent \rightarrow 1$
  as
  $\innovationprob \rightarrow 0$.
\end{textblock}

\begin{textblock}
\item
  However, it is evident that, in setting
  $\innovationprob = 0$ in Simon's model,
  the initial type at $\rgrtime=1$
  is replicated at every time step,
  leading to the size ranking
  of a system that contains only one type, i.e., 
\end{textblock}
\begin{equation}
  \typesize_{\rank,\rgrtime,\sizerankingexponent}
  =
  \left\{
  \begin{array}{l}
    \rgrtime \ \ \textnormal{if} \ \ \rank = 1, \\
    0 \ \ \textnormal{if} \ \ \rank > 1. \\
  \end{array}
  \right.
  \label{eq:rich-get-richer-complete.sizerank-zero-innovation}
\end{equation}

\begin{textblock}
\item
  Instead of $\sizerankingexponent = 1$,
  Eq.~\eqref{eq:rich-get-richer-complete.sizerank-zero-innovation} is the $\sizerankingexponent \rightarrow \infty$ limit.
\item
  Thus, the limiting size ranking of Simon's model for
  $\innovationprob \rightarrow 0$
  does not match that directly produced by starting with
  $\innovationprob = 0$.
\end{textblock}

\begin{textblock}
\item
  However, there is in fact no discontinuity in the solutions.
\item
  As documented in Ref.~\cite{dodds2017pre}, 
  Simon's model
  produces a hybrid size ranking
  with the lead type's size a factor
  $1/\innovationprob$
  greater
  than the next largest type, so that
  $
  \typesize_{1,\rgrtime,\sizerankingexponent}
  \propto
  \typesize_{2,\rgrtime,\sizerankingexponent}/\innovationprob
  $, i.e., a first-mover advantage effect. The tail behaves as in Eq.~\eqref{eq:rich-get-richer-complete.simons-tail}.
\end{textblock}

\section{Fundamental rich-get-richer mechanism}
\label{sec:rich-get-richer-complete.generalized-model}

\begin{textblock}
\item
  Here, we use mechanistic arguments to find a form of
  $\innovationprob_{\rgrtime,\sizerankingexponent}$
  which generalizes that of Simon's model
  to properly handle the
  $\innovationrate \rightarrow 0$
  limit, leaving the rest of the model unchanged.
\end{textblock}

\begin{textblock}
\item
  To do so, we return to an alternate
  treatment that directly estimates the size ranking,
  and avoids the outsized first-mover advantage. Let us focus on the $\rank$th type, $\typeinittime$ being the time at which the $\rank$th type appears. For a rich-get-richer process,
  the expected growth of the $\rank$th type
  at time $\rgrtime > \typeinittime$ is
\end{textblock}
\begin{equation}
  \left\langle
  \typesize_{\rank,\rgrtime+1,\sizerankingexponent}
  -
  \typesize_{\rank,\rgrtime,\sizerankingexponent}
  \right\rangle
  =
  (+1)
  (1-\innovationrate)
  \frac{
    \typesize_{\rank,\rgrtime,\sizerankingexponent}
  }{
    \rgrtime
  },
  \label{eq:rich-get-richer-complete.sizegrowth}
\end{equation}
\begin{textblock}
\item
  because the $\rank$th type already exists. Thus, the only way to grow is by refraining to innovate, hence the $(1-\innovationrate)$ term.
\end{textblock}

\begin{textblock}
\item 
  Given that the $\rank$th type first appears at initiation time $\typeinittime$ with size $\typesize_{\rank,\typeinittime,\sizerankingexponent}=1$, we then estimate its size at time $\rgrtime \ge \typeinittime$ as 
\item
  \begin{equation}
    \typesize_{\rank,\rgrtime,\sizerankingexponent}
    =
    \prod_{\dummytime = \typeinittime + 1}^{\rgrtime}
    \left[
      1
      +
      \frac{
        1-\innovationprob_{\dummytime}
      }{
        \dummytime
      }
      \right].
    \label{eq:rich-get-richer-complete.sizeestimate}
  \end{equation}
\item
  For $\rgrtime = \typeinittime$,
  we have the empty product 
  and
  $
  \typesize_{\rank,\typeinittime, \sizerankingexponent}
  =
  1
  $.

\end{textblock}\begin{textblock}
\item
  Let us consider two innovation rate regimes. First, to test the validity of the approach in Eq.~\eqref{eq:rich-get-richer-complete.sizeestimate}, we consider a regime with
  a fixed, non-zero $\innovationrate = \innovationprob$, not near 1. Second, we consider a size-dependent $\innovationprob$ evanating to zero
  as $\rgrtime \rightarrow \infty$, which differs from Simon's original constant innovation probability.
\end{textblock}

\begin{textblock}
\item
  For the first case, $0 < \innovationprob < 1$, we have
  $
  \typeinittime
  \sim
    \rank / \innovationprob
    ,
  $
\item
  and
  we retrieve the solution to Simon's model
  for
  $0 < \innovationprob \ll 1$,
  which agrees with the form
  of the idealized size ranking
  in Eq.~\eqref{eq:rich-get-richer-complete.simons-tail},
\end{textblock}
\begin{equation}
  \typesize_{\rank,\rgrtime,\sizerankingexponent}
  =
  \frac{
    B
    (
    \frac{
      \rank
    }{
      \innovationprob
    }
    +
    1,
    1
    -
    \innovationprob
    )
  }{
    B
    (
    \rgrtime
    +
    1,
    1
    -
    \innovationprob
    )
  }
  \sim
  \left[
    \frac{
      \innovationprob
      \rgrtime
    }{
      \rank
    }
    \right]^{1-\innovationprob}
  =
  \Ndistincttypes^{\sizerankingexponent}
  \rank^{-\sizerankingexponent},
  \label{eq:rich-get-richer-complete.sizefixedinnovation}
\end{equation}
\begin{textblock}
\item
  for
  $
  1
  \le
  \rank
  \le
  \Ndistincttypes
  $. 
  Here, we have identified the number of distinct types at time $\rgrtime$
  when
  $
  \innovationrate
  =
  \innovationprob
  $
  to be
  $
  \Ndistincttypes
  =
  \innovationprob
  \rgrtime
  $.
  
\item
  The size ranking's tail
  decays as an inverse power law with exponent
  $
  \sizerankingexponent
  =
  1
  -
  \innovationprob.
  $
  We thus recover Simon's original result for constant innovation rate,
\item 
  but only in the 
      $\innovationprob \ll 1$
  regime.
\end{textblock}

   \begin{textblock}
\item
  With the second case,
  where the innovation rate is dynamic, we want to fix the failure of Simon's model to describe $\sizerankingexponent \geq 1$. 
  If the $\innovationrate \rightarrow 0$
  fast enough as
  $\rgrtime \rightarrow \infty$ 
  (which we characterize properly below),
  then 
  Eq.~\eqref{eq:rich-get-richer-complete.sizeestimate}
  reduces to a telescoping product
  as
\item
  \begin{equation}
    \typesize_{\rank,\rgrtime,\sizerankingexponent}
    =
    \frac{
      \rgrtime
      +
      1
    }{
      \typeinittime
      +
      1
    }
    \sim
    \frac{
      \rgrtime
    }{
      \typeinittime
    }.
    \label{eq:rich-get-richer-complete.sizedecayinginnovation}
  \end{equation}
\item
  This implies that the tail of the size ranking behaves as the inverse of the initiation time of the $\rank$th type.
\item
  To obtain a power-law tail, we must evidently require that the initiation times grow as
  $
  \typeinittime
  \sim
  \rank^{\sizerankingexponent}
  /
  \rgrconstant$, 
  meaning
\item
  \begin{equation}
    \typesize_{\rank,\rgrtime,\sizerankingexponent}
    \sim
    \frac{
      \rgrconstant
      \,
      \rgrtime
      }{
      \rank^{\sizerankingexponent}
    },
    \label{eq:rich-get-richer-complete.sizedecayinginnovation-scaling}
  \end{equation}
\item
  where $\rgrconstant$ is a constant.
\item
  Thus for the rich-get richer mechanism to produce a size ranking with a power-law tail when $\sizerankingexponent > 1$, 
  the mechanism is power-law-in-power-law out (PLIPLO).
\item 
  While this may be considered less than satisfying, it is what is required for the basic rich-get-richer mechanism to deliver size rankings with $\innovationrate \rightarrow 0$ to explain $\sizerankingexponent \geq 1$.
\item 
  Our findings are concordant with that of Zanette and Montemurro
  on vocabulary growth for $\sizerankingexponent > 1$~\cite{Zanette2005}.
\end{textblock}

\begin{textblock}
\item
  Requiring the total type counts to be the time $\rgrtime$, we determine $\rgrconstant$ by simplifying the computations in the case of $\sizerankingexponent \gg 1$, so that
\begin{equation}
  \rgrtime
    = \sum_{\rank=1}^{\Ndistincttypes}
       \typesize_{\rank,\rgrtime,\sizerankingexponent}
       \simeq
       \rgrconstant \rgrtime
       \left( \frac{\Ndistincttypes^{1-\sizerankingexponent}}{1-\sizerankingexponent}+ \zeta(\sizerankingexponent) \right) 
       \sim \zeta(\sizerankingexponent) \rgrconstant \rgrtime
       \label{eq:rich-get-richer-complete.size-rank-norm-check-high-alpha}
\end{equation}
and $\rgrconstant = \zeta(\sizerankingexponent)^{-1}$, where we have used the Euler-Maclaurin expansion~\cite{apostol1999} of the harmonic sum.
\end{textblock}

\begin{textblock}
\item
  Equations~\eqref{eq:rich-get-richer-complete.sizedecayinginnovation}
  and
  \eqref{eq:rich-get-richer-complete.size-rank-norm-check-high-alpha}
  then give us that
  \begin{equation}   
  \typeinittime
  \sim
  \zeta(\sizerankingexponent)
  \rank^{\sizerankingexponent}.
\end{equation}
\item
  To determine
  $\innovationrate$, we use the fact that $\rank$ in the initialization time expression can be exchanged for
  $
  \Ndistincttypes
  $, 
  because at $\rgrtime = \typeinittime$ there will be $\Ndistincttypes = \rank$ types. Hence,
\item
  \begin{equation}
    \typeinittimefn{\Ndistincttypes}
    \sim 
    \zeta(\sizerankingexponent)
    \Ndistincttypes^{\sizerankingexponent},
    \label{eq:rich-get-richer-complete.tinit-scaling}
  \end{equation}
\item
  which is in accordance to the token-type scaling law for
  $\sizerankingexponent \gg 1$ also known as Heaps' law~\cite{rosillo-rodes2026a}.
\item
  From here, we can use the binomial expansion and estimate the innovation rate as
\item
  \begin{align}
  \notag
  \innovationrate
    &= \frac{1}{
        \typeinittimefn{\Ndistincttypes}
        - \typeinittimefn{\Ndistincttypes-1}
      } 
      \sim 
      \frac{1}{
        \zeta(\sizerankingexponent)\left[\Ndistincttypes^{\sizerankingexponent}
        - \left(\Ndistincttypes - 1\right)^{\sizerankingexponent}
      \right]} \\
      &
      \sim
      \frac{1}{\sizerankingexponent \zeta(\sizerankingexponent)}
      \Ndistincttypes^{1-\sizerankingexponent},
       \label{eq:rich-get-richer-complete.innovation-scaling-high-alpha}
\end{align}
\item
  which is consistent with $\sizerankingexponent \gg 1$ and $\innovationrate \rightarrow 0$.
\end{textblock}

\begin{textblock}
\item
  We are now able to construct a generalized innovation rate.
\item
  Summarizing, the desired limiting behavior of
  $\innovationrate$ is,
\item
  for $0 \leq \sizerankingexponent \ll 1$, 
  $
  \innovationrate
  \sim
  1 - \sizerankingexponent
  $, to recover the range in which Simon's framework works correctly,
\item 
  and for $\sizerankingexponent \gg 1$, 
  $
  \innovationrate
  \sim
  \left[\sizerankingexponent \zeta(\sizerankingexponent)\right]^{-1} \Ndistincttypes^{1-\sizerankingexponent}
  $.
\end{textblock}

\begin{textblock}
\item
  Focusing first on the behavior of $\innovationrate$ when $\sizerankingexponent \ll 1$, we can write
\item 
  \begin{equation}
    \innovationrate
    =
    \frac{
      1
      -
      \sizerankingexponent
    }{
      f(\Ndistincttypes,\sizerankingexponent)
    },
    \label{eq:rich-get-richer-complete.generalized-innovation-rate-proto-1}
  \end{equation}
\item
  where we must have, for bounded
  $\Ndistincttypes$,
  that 
  $
  f(\Ndistincttypes,\sizerankingexponent)
  \rightarrow
  1
  $
  when
  $
  \sizerankingexponent \ll 1
  $
  and 
  $
  f(\Ndistincttypes,\sizerankingexponent)
  \rightarrow
  \sizerankingexponent (1-\sizerankingexponent) \zeta(\sizerankingexponent) 
  \Ndistincttypes^{\sizerankingexponent-1}
  $
  when
  $
  \sizerankingexponent \gg 1
  $.
\item 
  The most parsimonious option is 
  \begin{equation}
      f(\Ndistincttypes,\sizerankingexponent)
  =
  1
  +
  \sizerankingexponent (1-\sizerankingexponent) \zeta(\sizerankingexponent)
  \Ndistincttypes^{\sizerankingexponent-1}.
  \end{equation}
\end{textblock}

\begin{textblock}
\item
  Our generalized
  innovation rate
  $\innovationrate$
  is then
\item 
  \begin{equation}
    \innovationrate
    =
    \frac{
      \dee \Ndistincttypes
    }{
      \dee \rgrtime
    }
    =
    \frac{
      1
      -
      \sizerankingexponent
    }{
      1
      +
      \sizerankingexponent (1-\sizerankingexponent) \zeta(\sizerankingexponent)
      (\Ndistincttypes+1)^{\sizerankingexponent - 1}
    }.
    \label{eq:rich-get-richer-complete.generalized-innovation-rate}
  \end{equation}
\end{textblock}  

\begin{textblock}
\item
  We have shifted 
  $\Ndistincttypes$
  to
  $\Ndistincttypes+1$
  so that the model functions sensibly in the first step. In this way,
\item
  given the initial condition of
  $
  \Ndistincttypesfn{0}{\sizerankingexponent}
  =
  0
  $,
  we have
  $\innovationratefn{1}{\sizerankingexponent} = 1$
  for all
  $\sizerankingexponent \ge 0$.
\item 
  The system will be seeded with an initial type
  at time
  $\rgrtime = 1$
  with
  $\Ndistincttypesfn{1}{\sizerankingexponent} = 1.$
\item
  For
  $
  \Ndistincttypes
  \rightarrow
  \infty
  $,
  the limiting form of
  the innovation rate is given by
\item
  \begin{equation}
    \innovationrate
    \rightarrow
    \left\{
    \begin{array}{cl}
      1 - \sizerankingexponent
      &
      \textnormal{for}
      \
      0 \le \sizerankingexponent \ll 1,
      \\
      \left(
      \ln \Ndistincttypes
        +
        \gamma_{0}
        \right)^{-1}
      &
      \textnormal{for}
      \
      \sizerankingexponent = 1,
      \\
      \left[\sizerankingexponent \zeta(\sizerankingexponent)\right]^{-1} \Ndistincttypes^{-(\sizerankingexponent-1)}
      &
      \textnormal{for}
      \
      \sizerankingexponent \gg 1,
    \end{array}
    \right.
    \label{eq:rich-get-richer-complete.innovation-rate-scaling-law}    
  \end{equation}
\item 
where $\gamma_{0} \simeq 0.577$ is the Euler-Mascheroni constant. 
\item 
See Appendix~\ref{app:limiting_behavior} for the derivation of the limits in Eq.~\eqref{eq:rich-get-richer-complete.innovation-rate-scaling-law} and Appendix~\ref{app:generalized_derivations} for derivations of Eqs.~\eqref{eq:rich-get-richer-complete.sizeestimate},~\eqref{eq:rich-get-richer-complete.sizefixedinnovation}, and~\eqref{eq:rich-get-richer-complete.sizedecayinginnovation}.
\item
Notice that the limiting form of $\innovationrate$ for $\sizerankingexponent \gg 1$ in Eq.~\eqref{eq:rich-get-richer-complete.innovation-rate-scaling-law} corresponds to Heaps' law~\cite{Zanette2005, rosillo-rodes2026a}.
\end{textblock}

\begin{textblock}
\item
Consequently for Zipf's law in the limit of large $\rgrtime$,
we find the `Zipf innovation rate':
\item 
\begin{equation}
  \innovationrate
  \rightarrow
  1 / \ln \Ndistincttypesfn{\rgrtime}{1}
  \label{eq:rich-get-richer-complete.zipf-innovation-rate}    
\end{equation}
\item 
Whereas Simon's analysis erroneously
had $\innovationprob \rightarrow 0$ for Zipf's law,
we find that the innovation rate must be non-zero
and decay more slowly than any inverse power law.
\end{textblock}

\begin{figure*}
\centering
 \includegraphics[width=\textwidth]{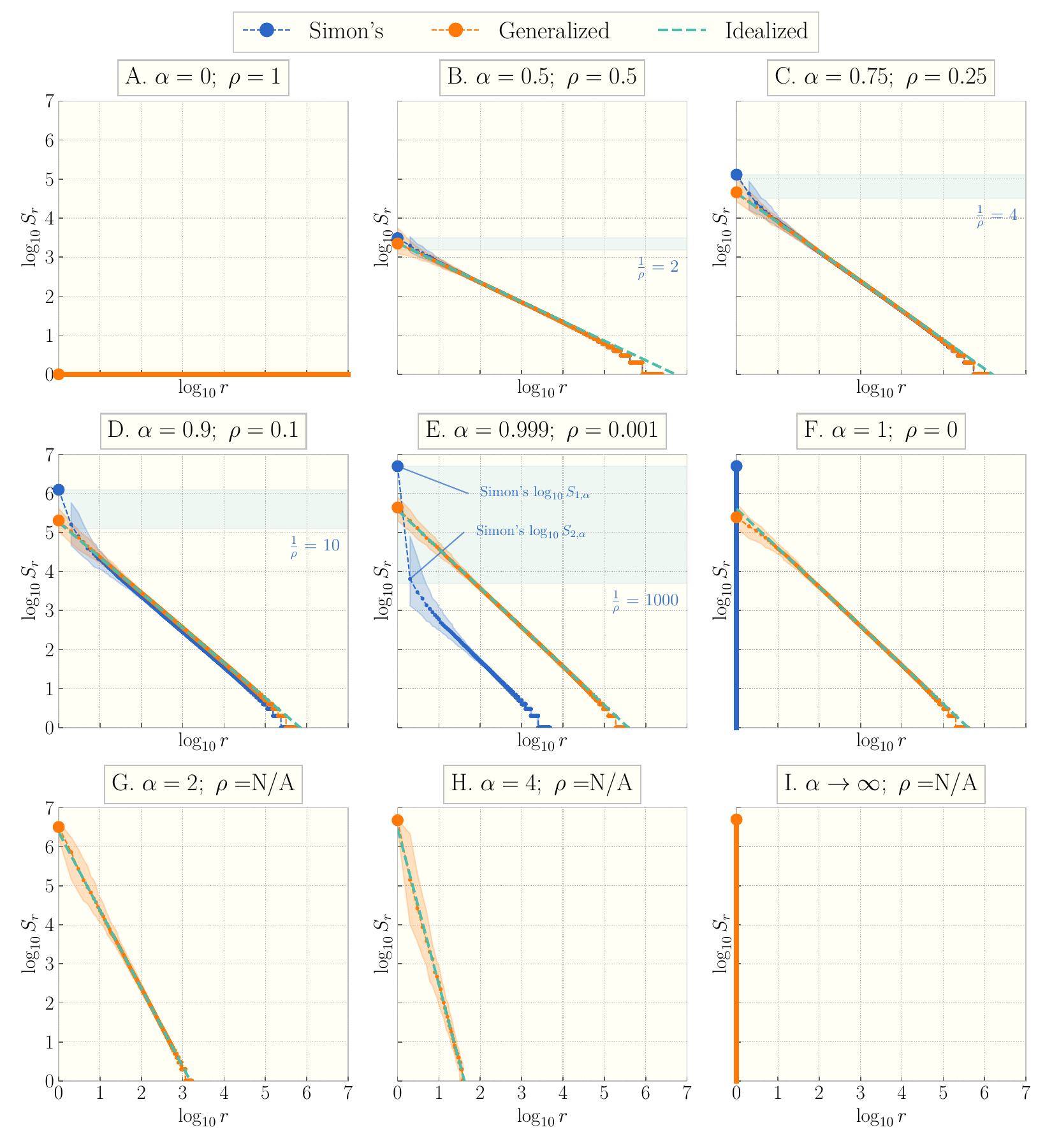}
 \caption{
 The generalized innovation rate $\innovationrate$ in Eq.~\eqref{eq:rich-get-richer-complete.generalized-innovation-rate} correctly reproduces the size–ranking $\typesize_{\rank,\rgrtime,\sizerankingexponent}$ for all values of $\sizerankingexponent$, including $\sizerankingexponent \geq 1$. Unlike Simon's model, which misrepresents the $\sizerankingexponent = 1 - \innovationprob = 1$ case by introducing an outsized first-mover advantage and shifting the size ranking toward $\sizerankingexponent \to \infty$, the generalized model we introduce in this work remains consistent across all regimes. We run 100 realizations of both models and show the median values and the 2.5 and 97.5 percentiles as colored shades. We also indicate with a shaded blue rectangle the amplitude of the first-mover advantage effect~\cite{dodds2017pre}.
 }
\label{fig:rich-get-richer-complete.rich_get_richer_complete_simulation100}
\end{figure*}

\begin{figure*}[t]
  \includegraphics[width=\linewidth]{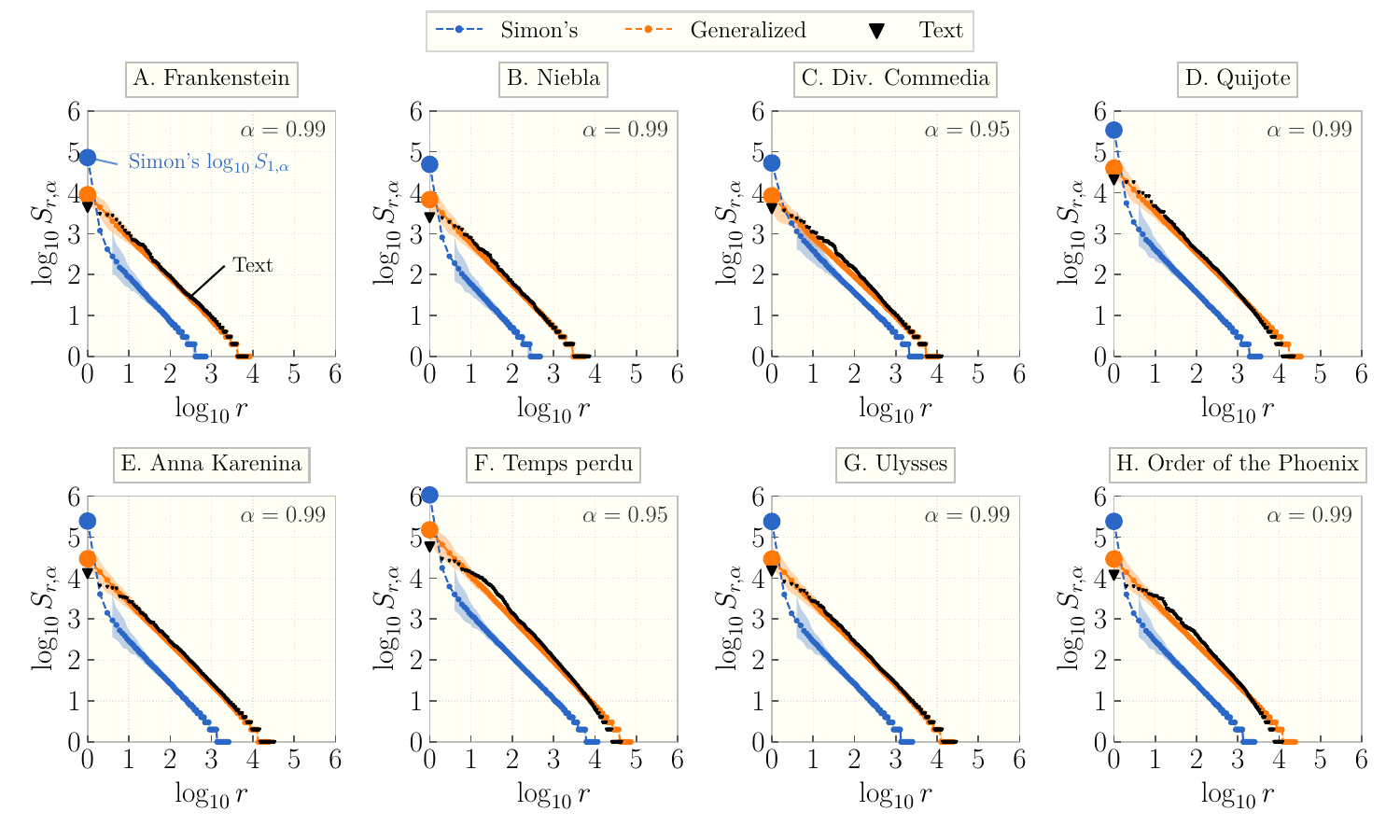}
  
  \caption{The generalized innovation rate defined in Eq.~\eqref{eq:rich-get-richer-complete.generalized-innovation-rate} accounts for real-world power-law behaviors such as those observed in size rankings of texts. We contrast Simon's original model with the generalized mechanism introduced here using data from five languages in eight different books: (\textbf{A}) \textit{Frankenstein} (English, 1818) by Mary Shelley; (\textbf{B}) \textit{Niebla} (Spanish, 1914) by Miguel de Unamuno; (\textbf{C}) \textit{La Divina Commedia} (Italian, 1320) by Dante Alighieri; (\textbf{D}) \textit{Don Quijote} (Spanish, 1605) by Miguel de Cervantes; (\textbf{E}) \textit{Anna Karenina} (Russian in Cyrillic alphabet, 1878) by Leo Tolstoy; (\textbf{F}) \textit{À la recherche du temps perdu} (French, 1913-1927) by Marcel Proust; (\textbf{G}) \textit{Ulysses} (English, 1922) by James Joyce; and (\textbf{H}) \textit{Harry Potter and the Order of the Phoenix} (English, 2003) by J. K. Rowling. We plot the median value of $\typesize_{\rank, \sizerankingexponent}$, and the colored areas bound 2.5 to 97.5 percentiles for $\typesize_{\rank, \sizerankingexponent}$ in 100 realizations of each model with $\sizerankingexponent = 0.99$, except for \textbf{C} and \textbf{F}, where $\sizerankingexponent = 0.95$. In all cases we kept $\sizerankingexponent < 1$ to allow for comparison between Simon's model and the generalized mechanism presented here. All the texts are drawn from the Standardized Project Gutenberg Corpus~\cite{Gerlach2020}, except for \textbf{E}, which is available in Lib.ru~\cite{libru}, \textbf{F}, which is sourced from La Bibliothèque électronique du Québec~\cite{beq}, and \textbf{H}, provided by the Computational Story Lab.
}
\label{fig:rich-get-richer-complete.books}
\end{figure*}

\begin{textblock}
\item
  In Fig.~\ref{fig:rich-get-richer-complete.rich_get_richer_complete_simulation100},
  we show results from simulations using
  the generalized $\innovationrate$
  from
  Eq.~\eqref{eq:rich-get-richer-complete.generalized-innovation-rate}
  for a range of
  $\sizerankingexponent$
  values. The generalized innovation rate $\innovationrate$ in Eq.~\eqref{eq:rich-get-richer-complete.generalized-innovation-rate} reproduces the intended rank–size profile $\typesize_{\rank,\rgrtime,\sizerankingexponent}$ across the full range of exponents $\sizerankingexponent$, including situations where the first-mover advantage is notable as in Fig.~\ref{fig:rich-get-richer-complete.rich_get_richer_complete_simulation100}\textbf{C}--\textbf{E}, and beyond the Zipf point, i.e., $\sizerankingexponent \ge 1$, as in Fig.~\ref{fig:rich-get-richer-complete.rich_get_richer_complete_simulation100}\textbf{F}--\textbf{I}. 
\end{textblock}

\begin{textblock}
  In Fig.~\ref{fig:rich-get-richer-complete.books},
  we demonstrate how our model reproduces the real-world size–rankings across eight novels across five languages.
\item 
  The generalized mechanism successfully accounts for the size rankings observed in all cases, whereas Simon's model does not. 
\end{textblock}  

\begin{textblock}
\item
    Importantly, we derived the generalized innovation rate in Eq.~\eqref{eq:rich-get-richer-complete.generalized-innovation-rate} by imposing a power-law form for the size ranking. 
    \item 
    In Appendix~\ref{app:inverse_rate_power-law} we show the opposite for $\sizerankingexponent=1$, that is, that substituting the asymptotic innovation rate into the size estimate, Eq.~\eqref{eq:rich-get-richer-complete.sizeestimate}, reproduces Zipf's law.
\end{textblock}

\section{Correspondence with growing, non-mechanistic families of size rankings}
\label{sec:rich-get-richer-complete.nonmech}

\begin{textblock}
\item
Here, we show that Eq.~\eqref{eq:rich-get-richer-complete.generalized-innovation-rate} 
is effectively produced
by a purely deterministic, growing model with a prescribed size ranking~\cite{rosillo-rodes2026a}.
\item 
For this model, the mechanism-free size ranking evolves as 
$
\typesize_{\rank,\rgrtime,\sizerankingexponent}
=
\left\lfloor
1/2
+
\growthfactor
\rank^{-\sizerankingexponent}
\right\rfloor$,
\item 
where
$\lfloor \cdot \rfloor$ is the floor operator
and $\growthfactor > 0$ is a tunable factor.
\item
We establish a one-to-one correspondence between the generalized innovation rate $\innovationrate$ in Eq.~\eqref{eq:rich-get-richer-complete.generalized-innovation-rate} and the type--token growth induced by this idealized model.
\end{textblock}

\begin{textblock}
\item
  Using $\rgrindex = 1, 2, \ldots$ as an index,
  we write
  \begin{equation}
    \bigprobsymbol_{\rank,\rgrindex,\sizerankingexponent}
    =
    \frac{1}{\rgrtime}
    \typesize_{\rank,\rgrindex,\sizerankingexponent}
    =
    \frac{
      \rank^{-\sizerankingexponent}
    }{
      \sum_{k=1}^{\Ndistincttypes}
      k^{-\sizerankingexponent}
    }
    ,
  \label{eq:rich-get-richer-complete.Pt}
  \end{equation}
\item
  where
  $1 \le \rank \le \Ndistincttypes$.
\item
  Now, $\Ndistincttypes$
  is determined by the indexed
  type-token relationship~\cite{rosillo-rodes2026a},
  \begin{equation}
    \rgrindex
        \sim
        \frac{
        \Ndistincttypesfn{\rgrindex}{\sizerankingexponent}
        }{
        1-\sizerankingexponent
        }
        +
        \zeta(\sizerankingexponent)
        \Ndistincttypesfn{\rgrindex}{\sizerankingexponent}^\sizerankingexponent.
    \label{eq:rich-get-richer-complete.type-token-index}
  \end{equation}
\item
  We determine how the number of distinct types changes
  with index $\rgrindex$ differentiating
  \Req{eq:rich-get-richer-complete.type-token-index}
  with respect to $\rgrindex$,
  so we have that
\item
  \begin{equation}
     \diff{\Ndistincttypesfn{\rgrindex}{\sizerankingexponent}}{\rgrindex}
    =
    \frac{
      1 - \sizerankingexponent
    }{
      1 + \sizerankingexponent (1-\sizerankingexponent) \zeta(\sizerankingexponent)
      \Ndistincttypesfn{\rgrindex}{\sizerankingexponent}^{\sizerankingexponent-1}
    }.
    \label{eq:rich-get-richer-complete.Ni-change}
  \end{equation}
\item
  We see that
  $\innovationrate$
  and
  $\diff{\Ndistincttypesfn{\rgrindex}{\sizerankingexponent}}{\rgrindex}$
  in Eqs.~\eqref{eq:rich-get-richer-complete.generalized-innovation-rate}
  and~\eqref{eq:rich-get-richer-complete.Ni-change}
  have the same limiting form, and are functionally equivalent.
\end{textblock}
\section{Concluding remarks}
\label{sec:rich-get-richer-complete.concludingremarks}

\begin{textblock}
\item
We have uncovered the minimal dynamic innovation rate that corrects and extends Simon's rich-get-richer model so as to produce power-law size rankings across all exponents, $0 \le \sizerankingexponent < \infty$.
\item
Having shown how Simon's model fails to actually produce Zipf's law 
in the limit of $\innovationprob \rightarrow 0$,
we determine the behavior of the dynamic innovation rate that does:
$\innovationratefn{\rgrtime}{1} 
\rightarrow 
1/ \log \Ndistincttypesfn{\rgrtime}{1}$.
\item 
Future work would explore the possible origins of this form.
\item
For novels, we have shown that our model accurately reproduces empirical size rankings while Simon's model does not.
\item
We have shown a deep congruence between our  
mechanistic rich–get–richer process and 
a general growing, non-mechanistic model.
\item
Any mechanism that leads to a growing power-law size ranking 
will functionally entail our generalized innovation rate.
\end{textblock}

\begin{textblock}
\item 
Our model is purposefully constrained by the assumption 
of a proportionate rich-get-richer process, which, although widespread, is not universal in empirical systems. 
\item
While our generalized innovation rate leads to the full range of power-law exponents, additional mechanisms known to generate power law size rankings could further expand the framework's scope and realism.
\item
Finally, for $\sizerankingexponent > 1$, the power-law-in–power-law-out nature of the mechanism invites explication of an underlying process that could give rise to a power-law decaying innovation rate.
\item
These limitations aside, our generalized innovation rate provides a coherent reference for understanding power law size rankings,
repairs and goes beyond Simon's model,
and will serve as a base model for all rich-get-richer systems.
\end{textblock}



\section*{Acknowledgments}
The authors thank Santiago Lamata-Otín, Lute Lillo, and Gema Mora for useful discussions.
P.R. acknowledges support by the Spanish State Research Agency (MICIU/AEI/10.13039/501100011033) and FEDER (UE) under the Mar{\'\i}a de Maeztu project CEX2021-001164-M and
the projects PID2021-122256NB-C21 and PID2024-157493NB-C21, during a research stay at the Computational Story Lab.
We are grateful for
National Science Foundation Award \#2242829
(Science of Online Corpora, Knowledge, and Stories) and award \#2419733,
foundational support from MassMutual,
and
an anonymous philanthropic gift. The Vermont Advanced Computing Center (VACC) at the University of Vermont provided computational resources that contributed to the research results reported in this paper.




\bibliography{\filenamebase}

\begin{thebibliography}{10}

\bibitem{estoup1916a}
J.~B. Estoup.
\newblock {\em Gammes sténographiques}.
\newblock Institut Stenographique de France, Paris, 1916.

\bibitem{zipf1946}
G.~K. Zipf.
\newblock {The psychology of language}.
\newblock In {\em Encyclopedia of psychology}, pages 332--341. Philosophical Library, 1946.

\bibitem{altmann2016}
E.~G. Altmann and M.~Gerlach.
\newblock {\em {Statistical Laws in Linguistics}}, pages 7--26.
\newblock Springer International Publishing, Cham, 2016.

\bibitem{arnon2025}
I.~Arnon, S.~Kirby, J.~A. Allen, C.~Garrigue, E.~L. Carroll, and E.~C. Garland.
\newblock Whale song shows language-like statistical structure.
\newblock {\em Science}, 387(6734):649--653, 2025.

\bibitem{camacho2001}
J.~Camacho and R.~V. Solé.
\newblock Scaling in ecological size spectra.
\newblock {\em Europhysics Letters}, 55(6):774, sep 2001.

\bibitem{serrano2003}
M.~A. Serrano and M.~Bogu\~n\'a.
\newblock Topology of the world trade web.
\newblock {\em Phys. Rev. E}, 68:015101, Jul 2003.

\bibitem{gabaix2016}
X.~Gabaix.
\newblock {Power laws in economics: An introduction}.
\newblock {\em Journal of Economic Perspectives}, 30(1):185--206, 2016.

\bibitem{price1965a}
D.~J. de~Solla~Price.
\newblock Networks of scientific papers.
\newblock {\em Science}, 149:510--515, 1965.

\bibitem{Redner1998}
S.~Redner.
\newblock {How popular is your paper? An empirical study of the citation distribution}.
\newblock {\em The European Physical Journal B}, 4:131--134, July 1998.

\bibitem{mitzenmacher2004}
M.~Mitzenmacher.
\newblock {A brief history of generative models for power law and lognormal distributions}.
\newblock {\em Internet mathematics}, 1(2):226--251, 2004.

\bibitem{newman2005}
M.~E. Newman.
\newblock {Power laws, Pareto distributions and Zipf's law}.
\newblock {\em Contemporary physics}, 46(5):323--351, 2005.

\bibitem{maillart2008a}
T.~Maillart, D.~Sornette, S.~Spaeth, and G.~von Krogh.
\newblock Empirical tests of {Z}ipf's law mechanism in open source {L}inux distribution.
\newblock {\em Phys. Rev. Lett.}, 101(21):218701, 2008.

\bibitem{roman2022}
S.~Roman and F.~Bertolotti.
\newblock {A master equation for power laws}.
\newblock {\em Royal Society open science}, 9(12):220531, 2022.

\bibitem{holehouse2025}
J.~Holehouse, S.~Redner, V.~C. Yang, P.~L. Krapivsky, J.~I. Arroyo, G.~B. West, C.~Kempes, and H.~Youn.
\newblock A generative model of function growth explains hidden self-similarities across biological and social systems.
\newblock {\em arXiv preprint arXiv:2509.14468}, 2025.

\bibitem{ferrericancho2010a}
R.~Ferrer-i\mbox{-}Cancho and B.~Elvev\r{a}g.
\newblock Random texts do not exhibit the real {Z}ipf's law-like rank distribution.
\newblock {\em PLoS ONE}, 5:e9411, 03 2010.

\bibitem{broido2019}
A.~D. Broido and A.~Clauset.
\newblock Scale-free networks are rare.
\newblock {\em Nature Communications}, 10(1):1017, 2019.

\bibitem{gerlach2019}
M.~Gerlach and E.~G. Altmann.
\newblock {Testing Statistical Laws in Complex Systems}.
\newblock {\em Phys. Rev. Lett.}, 122:168301, Apr 2019.

\bibitem{simon1955a}
H.~A. Simon.
\newblock On a class of skew distribution functions.
\newblock {\em Biometrika}, 42:425--440, 1955.

\bibitem{yule1925a}
G.~U. Yule.
\newblock A mathematical theory of evolution, based on the conclusions of {D}r {J}. {C}. {W}illis, {F}.{R}.{S}.
\newblock {\em Phil. Trans. B}, 213:21--87, 1925.

\bibitem{merton1988matthewII}
R.~K. Merton.
\newblock {The Matthew Effect in Science, II: Cumulative Advantage and the Symbolism of Intellectual Property}.
\newblock {\em Isis}, 79(4):606--623, 1988.

\bibitem{albert1999}
R.~Albert, H.~Jeong, and A.-L. Barabási.
\newblock Diameter of the world-wide web.
\newblock {\em Nature}, 401(6749):130--131, 1999.

\bibitem{mandelbrot1959a}
B.~B. Mandelbrot.
\newblock A note on a class of skew distribution function. {A}nalysis and critique of a paper by {H}.~{A}.~{S}imon.
\newblock {\em Information and Control}, 2:90--99, 1959.

\bibitem{simon1960a}
H.~A. Simon.
\newblock Some further notes on a class of skew distribution functions.
\newblock {\em Information and Control}, 3:80--88, 1960.

\bibitem{mandelbrot1961a}
B.~B. Mandelbrot.
\newblock Final note on a class of skew distribution functions: analysis and critique of a model due to {H}. {A}. {S}imon.
\newblock {\em Information and Control}, 4:198--216, 1961.

\bibitem{simon1961a}
H.~A. Simon.
\newblock Reply to `final note' by {B}eno\^{i}t {M}andelbrot.
\newblock {\em Information and Control}, 4:217--223, 1961.

\bibitem{mandelbrot1961b}
B.~B. Mandelbrot.
\newblock Post scriptum to `final note'.
\newblock {\em Information and Control}, 4:300--304, 1961.

\bibitem{simon1961b}
H.~A. Simon.
\newblock Reply to {D}r.~{M}andelbrot's post scriptum.
\newblock {\em Information and Control}, 4:305--308, 1961.

\bibitem{dodds2017pre}
P.~S. Dodds, D.~R. Dewhurst, F.~F. Hazlehurst, C.~M. Van~Oort, L.~Mitchell, A.~J. Reagan, J.~R. Williams, and C.~M. Danforth.
\newblock Simon's fundamental rich-get-richer model entails a dominant first-mover advantage.
\newblock {\em Phys. Rev. E}, 95:052301, May 2017.

\bibitem{Zanette2005}
D.~Zanette and M.~Montemurro.
\newblock {Dynamics of Text Generation with Realistic Zipf's Distribution}.
\newblock {\em Journal of Quantitative Linguistics}, 12(1):29--40, 2005.

\bibitem{rosillo-rodes2026a}
P.~Rosillo-Rodes, L.~H\'{e}bert-Dufresne, and P.~S. Dodds.
\newblock Complete asymptotic type-token relationship for growing complex systems with inverse power-law count rankings.
\newblock {\em Physical Review Research}, 8(1):L012029, 2026.

\bibitem{apostol1999}
T.~M. Apostol.
\newblock {An Elementary View of Euler's Summation Formula}.
\newblock {\em The American Mathematical Monthly}, 106(5):409--418, 1999.

\bibitem{Gerlach2020}
M.~Gerlach and F.~Font-Clos.
\newblock {A Standardized Project Gutenberg Corpus for Statistical Analysis of Natural Language and Quantitative Linguistics}.
\newblock {\em Entropy}, 22(1):126, 2020.

\bibitem{libru}
Lib.ru: Klassika.
\newblock \url{http://az.lib.ru/}, 2004.
\newblock Supported by the Federal Agency for Press and Mass Communications.

\bibitem{beq}
{La Bibliothèque électronique du Québec}.
\newblock \url{http://beq.ebooksgratuits.com}.
\newblock Textes d'auteurs du domaine public. Fondée en 1998.

\end{thebibliography}




\clearpage

\appendix

\section*{Appendix}

\setcounter{page}{1}
\renewcommand{\thepage}{A\arabic{page}}
\renewcommand{\thefigure}{A\arabic{figure}}
\renewcommand{\thetable}{A\arabic{table}}
\renewcommand{\theequation}{A\arabic{equation}}
\setcounter{figure}{0}
\setcounter{table}{0}
\setcounter{equation}{0}

\renewcommand{\thesection}{A\arabic{section}}
\setcounter{section}{0}

\section{Limiting behavior of the generalized innovation rate}
\label{app:limiting_behavior}

\begin{textblock}
\item   
    Here, we derive the generalized innovation rate limits in Eq.~\ref{eq:rich-get-richer-complete.innovation-rate-scaling-law}.
\end{textblock}

\begin{textblock}
\item 
    In the case of $0 < \sizerankingexponent \ll 1$,
    \begin{equation}
        \innovationrate 
        \sim
        \frac{
        1-\sizerankingexponent
        }{
        1-\frac{
        \sizerankingexponent
        }{
        \Ndistincttypes + 1
        }
        }
        \simeq
        1 - \alpha.
    \end{equation}
\item 
    When $\sizerankingexponent \gg 1$,
    \begin{equation}
        \innovationrate
        \sim
        \left(
        \Ndistincttypes + 1
        \right)^{1-\sizerankingexponent}
        \simeq
        \Ndistincttypes.
    \end{equation}

\item 
    Finally, for deriving the $\sizerankingexponent \rightarrow 1$ limit, let us set $\sizerankingexponent = 1-\delta$, $|\delta| \ll 1$. Then,
    \begin{equation}
        \left( \Ndistincttypes + 1 \right)^{-\delta} 
        =
        e^{-\delta \ln \left( \Ndistincttypes + 1 \right)}
        \simeq
        1
        -
        \delta \ln \left( \Ndistincttypes + 1 \right)
    \end{equation}
    and
    \begin{equation}
        \innovationrate
        \simeq
        \frac{
        \delta
        }{
        1
        -
        \left[
        1 - \delta - \delta \ln \left(\Ndistincttypes + 1\right)
        \right]
        }
        =
        \frac{1
        }{
        1 + \ln \left(
        \Ndistincttypes + 1
        \right)
        }.
    \end{equation}
\end{textblock}

\section{Detailed derivations of the generalized rich-get-richer mechanism }
\label{app:generalized_derivations}

Eq.~\eqref{eq:rich-get-richer-complete.sizeestimate} follows from Eq.~\eqref{eq:rich-get-richer-complete.sizegrowth}, so that after dropping expectations,
\begin{equation}
\typesize_{\rank,\rgrtime,\sizerankingexponent}
    =
\left[
    1
    +
    \frac{
    1
    -
    \innovationratefn{\rgrtime - 1}{\sizerankingexponent}
    }{
    \rgrtime - 1
    }
\right]
\typesize_{\rank,\rgrtime - 1,\sizerankingexponent}.
\end{equation}
Leveraging the mean-field estimation that a trajectory follows its expectation, we have that
\begin{align}
    \typesize_{\rank,\rgrtime,\sizerankingexponent} 
    &
    = 
    \left[
    1
    +
    \frac{
    1
    -
    \innovationratefn{\rgrtime - 1}{\sizerankingexponent}
    }{
    \rgrtime - 1
    }
\right]
 \typesize_{\rank,\rgrtime - 1,\sizerankingexponent},
\nonumber 
\\
&
=
    \left[
    1
    +
    \frac{
    1
    -
    \innovationratefn{\rgrtime - 2}{\sizerankingexponent}
    }{
    \rgrtime - 2
    }
\right]
    \left[
    1
    +
    \frac{
    1
    -
    \innovationratefn{\rgrtime - 1}{\sizerankingexponent}
    }{
    \rgrtime - 1
    }
\right]
\typesize_{\rank,\rgrtime - 2,\sizerankingexponent},
\nonumber
\\
&
=
\ldots
=
\prod_{\dummytime = \typeinittime + 1}^{\rgrtime}
    \left[
      1
      +
      \frac{
        1-\innovationprob_{\dummytime}
      }{
        \dummytime
      }
      \right]
      \cdot 
      1,
\end{align}
where we have used that the $\rank$th type appears at $\typeinittime$ with size $\typesize_{\rank,\typeinittime,\sizerankingexponent} = 1$. We will take this result to derive Eq.~\eqref{eq:rich-get-richer-complete.sizefixedinnovation}. Given a constant innovation rate such that $0 < \innovationprob \ll 1$, we have that 
\begin{align}
    \typesize_{\rank,\rgrtime,\sizerankingexponent} 
    &
    = 
    \prod_{\dummytime = \typeinittime + 1}^{\rgrtime}
    \left[
      1
      +
      \frac{
        1-\innovationprob
      }{
        \dummytime
      }
      \right]
      \nonumber 
      \\
      &
      =
      \prod_{\dummytime = \typeinittime + 1}^{\rgrtime}
    \left[
      \frac{
        \dummytime + 1-\innovationprob
      }{
        \dummytime
      }
      \right]
      \nonumber 
      \\  
      &
      =
      \frac{
      \Gamma\left(\typeinittime + 1\right) \Gamma\left(1-\innovationprob+\rgrtime+1\right)
      }{
      \Gamma\left(\rgrtime+1\right) \Gamma\left(\typeinittime+1+1-\innovationprob\right)
      }
      \nonumber 
      \\
      & 
      =
      \frac{      
      \Gamma\left(\typeinittime+1\right)
      }{
      \Gamma\left(\rgrtime+1\right)
      }
      \frac{
      \Gamma\left(\rgrtime+2-\innovationprob\right)
      }{
      \Gamma\left(\typeinittime+2-\innovationprob\right)
      }
      \nonumber 
      \\
      &
      =
      \frac{
      B\left(\typeinittime+1,1-\innovationprob\right)
      }{
      B\left(\rgrtime+1,1-\innovationprob\right)
      },
\end{align}
where we have used $ y = 1-\innovationprob$ in the identity
\begin{equation}
    B(x,y) = \frac{
    \Gamma(x) \Gamma(y)
    }{
    \Gamma(x+y)
    }.
\end{equation}
\begin{textblock}
\item 
As $\innovationprob$ is constant, $\typeinittime \sim \rank/\innovationprob$. 
\item
Also, because 
for fixed $y$ and large $x$,
$B(x,y) 
\rightarrow 
\Gamma(y) 
x^{-y}$, 
we have that    
\end{textblock}
\begin{equation}
    \typesize_{\rank,\rgrtime,\sizerankingexponent}
    \sim
    \frac{
    \Gamma(1-\innovationprob) \left(\frac{\rank}{\innovationprob}\right)^{\innovationprob-1}
    }{
    \Gamma(1-\innovationprob) \left(\rgrtime+1\right)^{\innovationprob-1}
    } 
    \simeq 
    \rank^{\innovationprob-1} 
    (\innovationprob \rgrtime)^{1-\innovationprob}.
\end{equation}
As $\Ndistincttypes = \innovationprob \rgrtime$ when $\innovationprob$ is constant, then
\begin{equation}
    \typesize_{\rank,\rgrtime,\sizerankingexponent} 
    \sim 
    \Ndistincttypes^\sizerankingexponent
    \rank^{-\sizerankingexponent},
\end{equation}
where 
$\sizerankingexponent
=
1 - \innovationprob.$

\section{Inverse reasoning for \texorpdfstring{$\sizerankingexponent = 1$}{the case of Zipf's law}} 
\label{app:inverse_rate_power-law}

By imposing a power-law behaviour in Eq.~\eqref{eq:rich-get-richer-complete.sizeestimate}, we arrive in Section~\ref{sec:rich-get-richer-complete.generalized-model} to the generalized innovation rate in Eq.~\eqref{eq:rich-get-richer-complete.generalized-innovation-rate}. 

Here, we show that, for the case $\sizerankingexponent = 1$, we can actually plug in the limiting behavior of Eq.~\eqref{eq:rich-get-richer-complete.generalized-innovation-rate} into Eq.~\eqref{eq:rich-get-richer-complete.sizeestimate} and arrive to Zipf's law without imposing any specific shape for $\typesize_{\rank,\rgrtime,\sizerankingexponent}$.

Let us begin with
\begin{equation}
    \innovationprob_{\rgrtime,1} 
    = 
    \frac{
    \mathrm{d} \Ndistincttypesfn{\rgrtime}{1}
    }{
    \mathrm{d}\rgrtime
    }
    =
    \frac{
    1
    }{
    1 + \ln \left( \Ndistincttypesfn{\rgrtime}{1} + 1 \right)
    }.
    \label{appeq:dNdt}
\end{equation}

Integrating Eq.~\eqref{appeq:dNdt}, we get $\rgrtime \sim \Ndistincttypesfn{\rgrtime}{1} \ln \Ndistincttypesfn{\rgrtime}{1}$, so 
\begin{equation}
    \typeinittime \sim r \ln r
\end{equation}
and
\begin{equation}
    \typesize_{r,\rgrtime} = \frac{
    \rgrtime+1
    }{
    \typeinittime + 1
    }
    \prod_{\dummytime = \typeinittime + 1}^{\rgrtime}
        \left(
        1
        -
        \frac{
        \innovationprob_{\dummytime,1}
        }{
        \dummytime + 1
        }
        \right).
\end{equation}

Let $\tau_{n}$ be the time at which $\Ndistincttypesfn{\tau_{n}}{1} = n$. Then,  $\forall \dummytime \in [\tau_{n}, \tau_{n+1})$, $\Ndistincttypesfn{\dummytime}{1} = n$, so $\innovationprob_{\tau_{n}, 1} \equiv \innovationprob_{n,1}$ is a constant. Then, let us define
\begin{equation}
    \mathcal{P}
    \equiv
    \prod_{n = r}^{\Ndistincttypesfn{\rgrtime}{1} - 1}
    \prod_{\dummytime = \tau_{n}}^{\tau_{n+1}-1}
    \left(
    1
    -
    \frac{
    \innovationprob_{n,1}
    }{
    \dummytime+1
    }
    \right),
\end{equation}
where we have used that $\tau_{r} = \typeinittime$. Taking $\ln \mathcal{P}$, we can use that $\ln(1-x) \sim -x$ because $\innovationprob_{n,1}$ decays sufficiently fast with $\rgrtime$, i.e., with $n$. That leaves us with
\begin{equation}
    \ln \mathcal{P}
    \sim
    -
    \sum_{n = r}^{\Ndistincttypesfn{\rgrtime}{1} - 1}
    \sum_{\dummytime = \tau_{n}}^{\tau_{n+1}-1}
    \frac{
    \innovationprob_{n,1}
    }{
    \dummytime+1
    }.
\end{equation}

We now deal with the sum in $\dummytime$. Using the asymptotic form of the harmonic sum~\cite{apostol1999}, if
\begin{equation}
    H_{k} = \sum_{i=1}^{k} i^{-1} \sim \ln k,
\end{equation}
then
\begin{align}
\notag
    \sum_{\dummytime = \tau_{n}}^{\tau_{n+1}-1}
    \frac{
    \innovationprob_{n,1}
    }{
    \dummytime+1
    }
    &=
    \sum_{\dummytime = \tau_{n}+1}^{\tau_{n+1}}
    \frac{
    \innovationprob_{n,1}
    }{
    \dummytime
    }
    =
    H_{\tau_{n+1}} - H_{\tau_{n}} 
    \\
    &\sim
    \ln \left(
        1
        +
        \frac{
            \varepsilon
        }{
            \tau_{n}
        }
        \right),
\end{align}
with $\varepsilon = \tau_{n+1} - \tau_{n}$.

We have that $\langle \varepsilon \rangle = \innovationprob_{n,1}^{-1}$, so
\begin{equation}
    \left\langle 
        \sum_{\dummytime = \tau_{n}}^{\tau_{n+1}-1}
        \frac{
        \innovationprob_{n,1}
        }{
        \dummytime+1
        }
    \right\rangle
    \sim
    \ln \left(
        1
        +
        \frac{
        1
        }{
        \innovationprob_{n,1} \tau_{n}
        }
    \right).
\end{equation}
Knowing that $\innovationprob_{n,1}^{-1}$ grows as $1+\ln n$ and $\tau_{n}$ as $n \ln n$ because of the integration of Eq.~\eqref{appeq:dNdt}, we see that $\tau_{n}$ grows faster than $\innovationprob_{n,1}^{-1}$, so $\innovationprob_{n,1}^{-1}$ $\tau_{n}^{-1}$ is small and 
\begin{equation}
    \ln \mathcal{P}
    =
    -
    \sum_{n = r}^{\Ndistincttypesfn{\rgrtime}{1} - 1}
    \innovationprob_{n,1} \innovationprob_{n,1}^{-1} \tau_{n}^{-1}
    \sim
    \sum_{n = r}^{\Ndistincttypesfn{\rgrtime}{1} - 1}
    \frac{
    1
    }{
    n \ln n
    }.
    \label{appeq:summation_nlnn}
\end{equation}
The argument of the sum in Eq.~\eqref{appeq:summation_nlnn} is, for $n > e$, monotonically decreasing, smooth and positive. We can limit the sum with two integrals, so that
\begin{equation}
    \int_{r}^{\Ndistincttypesfn{\rgrtime}{1}} 
        \frac{
            \mathrm{d}k
            }{
            k \ln k
            }
    \leq
    \sum_{k = r}^{\Ndistincttypesfn{\rgrtime}{1} - 1}
        \frac{
        1
        }{
        k \ln k
        }
    \leq
    \int_{r-1}^{\Ndistincttypesfn{\rgrtime}{1}-1} 
        \frac{
            \mathrm{d}k
            }{
            k \ln k
            },
\end{equation}
and then
\begin{align}
\notag
    \ln \mathcal{P}
    &\sim
    - \ln \ln \Ndistincttypesfn{\rgrtime}{1}
    - \ln \ln r
    \\
    &+
    \mathcal{O}
    \left[
    \left(r \ln r\right)^{-1} + \left(\Ndistincttypesfn{\rgrtime}{1} \ln \Ndistincttypesfn{\rgrtime}{1} \right)^{-1}
    \right].
\end{align}
With all this, we arrive at
\begin{equation}
    \typesize_{r,\rgrtime} \sim r^{-1}.
\end{equation}

\begin{figure}[t]
\centering
  \includegraphics[width=1\columnwidth]{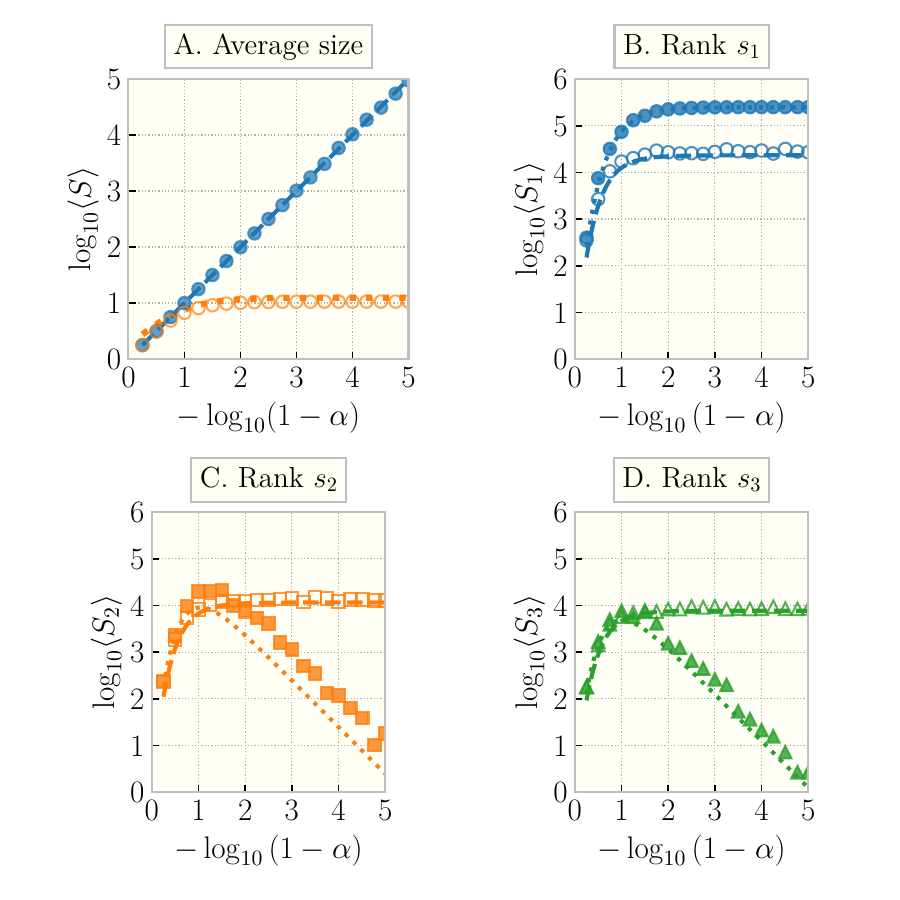}
  \caption{\textbf{A.} 
  Comparison of the behavior of average type size $\langle \typesize \rangle$ 
  as $\sizerankingexponent \rightarrow 1^{-}$
  for Simon's model and our generalized model.
  The curves are for $\rgrtime = 2.5\times10^{5}$ for Simon's model (filled) and our proposed generalized mechanism (empty)
  corresponding to Eqs.~\eqref{eq:avgsizesimons} (dashed line) and~\eqref{eq:avgsizegeneralized} (dotted line). 
  \textbf{B, C, and D.} $\langle \typesize_{1} \rangle$ (circles), $\langle \typesize_{2} \rangle$ (squares), and $\langle \typesize_{3} \rangle$ (triangles) at $\rgrtime = 2.5\times 10^{5}$ for Simon's model (filled) and our proposed mechanism (empty), representing the analytical expansions in Eqs.~\eqref{eq:siscalesimons} (dotted line) and~\eqref{eq:siscalegen} (dashed line). All the results correspond to averaging over 25 realizations.
  }
  \label{fig:rich-get-richer-complete.size\typesize_average_combined}
\end{figure}

\section{Analysis of the difference between Simon's and the generalized models using \texorpdfstring{$\langle \typesize \rangle$ and $\typesize_{i}$}{average type size and type size by rank}}

For large $\rgrtime$ and $\sizerankingexponent = 1-\varepsilon$ we have that 
\begin{equation}
    \langle 
    \typesize 
    \rangle^{\mathrm{Simon}} 
    = 
    \frac{
      \rgrtime
    }{
       \Ndistincttypesfn{\rgrtime}{\innovationprob}^{\mathrm{Simon}}
    } 
    = 
    \frac{
      \rgrtime
    }{
      1+\innovationprob (\rgrtime -1)
    } 
    \sim 
    \frac{1}{\varepsilon},
    \label{eq:avgsizesimons}
\end{equation}
with $\Ndistincttypesfn{\rgrtime}{\innovationprob}^{\mathrm{Simon}}$ 
being the 
expected number of unique types or groups at time $\rgrtime$ with an innovation rate $\innovationprob$ for Simon's model.

For a growing 
non-mechanistic
system (NMS)
with a stable power-law growth, we know from Ref.~\cite{rosillo-rodes2026a} that
\begin{align}
    \notag
    \rgrtime &\sim \frac{1}{1-\sizerankingexponent} \Ndistincttypesfn{\rgrtime}{\sizerankingexponent} + \zeta(\sizerankingexponent)\Ndistincttypesfn{\rgrtime}{\sizerankingexponent}^{\sizerankingexponent} \\
    \notag
    &\sim \Ndistincttypesfn{\rgrtime}{\sizerankingexponent} \ln \Ndistincttypesfn{\rgrtime}{\sizerankingexponent} - \frac{\Ndistincttypesfn{\rgrtime}{\sizerankingexponent}}{2} \varepsilon \ln^{2} \Ndistincttypesfn{\rgrtime}{\sizerankingexponent} + \Ndistincttypesfn{\rgrtime}{\sizerankingexponent} \gamma_{0} \\
    &- \Ndistincttypesfn{\rgrtime}{\sizerankingexponent} \varepsilon \gamma_{0} \ln \Ndistincttypesfn{\rgrtime}{\sizerankingexponent} + \Ndistincttypesfn{\rgrtime}{\sizerankingexponent} \gamma_{1} \varepsilon,
\end{align}
$\gamma_{i}$ being the $i$-th Stieltjes constant. To leading order, $\rgrtime\sim \Ndistincttypesfn{\rgrtime}{\sizerankingexponent} \ln \Ndistincttypesfn{\rgrtime}{\sizerankingexponent}$ so, using the series expansion of Lambert's W, we have that $\Ndistincttypesfn{\rgrtime}{\sizerankingexponent} \sim \rgrtime / \ln \rgrtime$ to leading order.

To solve $\Ndistincttypesfn{\rgrtime}{\sizerankingexponent}$ we use the ansatz
\begin{equation}
    \ln \Ndistincttypesfn{\rgrtime}{\sizerankingexponent} \sim \ln \rgrtime - \ln \ln \rgrtime + \delta,
\end{equation}
$\delta$ being a small perturbation to the leading order behavior. Solving for $\delta$ we get
\begin{equation}
    \Ndistincttypesfn{\rgrtime}{\sizerankingexponent} \sim \frac{\rgrtime}{\ln \rgrtime}\left[1 + \frac{\varepsilon}{2} \ln \rgrtime + \frac{\ln \ln \rgrtime - \left( \gamma_{0} + \varepsilon \gamma_{1} \right)}{\ln \rgrtime}\right],
\end{equation}
which leads to
\begin{align}
    \notag
    \langle \typesize \rangle^{\mathrm{NMS}} &= \frac{\ln \rgrtime}{1+\frac{\varepsilon}{2} \ln \rgrtime + \frac{\ln \ln \rgrtime - \left( \gamma_{0} + \varepsilon \gamma_{1} \right)}{\ln \rgrtime}} \\
    &\sim \frac{\ln \rgrtime}{1+\frac{\varepsilon}{2} \ln \rgrtime }.
    \label{eq:avgsizegeneralized}
\end{align}

In Fig.~\ref{fig:rich-get-richer-complete.size\typesize_average_combined}\textbf{A},
we compare the behavior of average type size for
Simon's model and our generalized model
as 
$\sizerankingexponent \rightarrow 1^{-}.$

Regarding $\typesize_{i}$, we know from Ref.~\cite{dodds2017pre} that 
\begin{equation}
    \typesize_{\rank,\rgrtime}^{\mathrm{Simon}} \sim \begin{cases} 
\frac{1}{\Gamma(2-\varepsilon)} \left[ \frac{1}{\rgrtime} \right]^{-(1-\varepsilon)} & \text{for } \rank = 1, \\ 
\varepsilon^{1-\varepsilon} \left[ \frac{\rank-1}{\rgrtime} \right]^{-(1-\varepsilon)} & \text{for } \rank \geq 2, 
\end{cases}
\label{eq:siscalesimons}
\end{equation}
and from Ref.~\cite{rosillo-rodes2026a}, that
\begin{equation}
    \typesize_{\rank,\rgrtime}^{\mathrm{NMS}} \sim \Ndistincttypesfn{\rgrtime}{\sizerankingexponent}^{\sizerankingexponent} \rank^{-\sizerankingexponent}.
    \label{eq:siscalegen}
\end{equation}
We compare their behavior in 
Fig.~\ref{fig:rich-get-richer-complete.size\typesize_average_combined}\textbf{B}-\textbf{D}.


\end{document}